\documentstyle[12pt]{article}

\def\Lheader #1{\gdef\LLheader{#1}}
\gdef\LLheader{}
\def\Ldate #1{\gdef\LLdate{#1}}
\gdef\LLdate{\today}
\def\Ltitle #1{\gdef\LLtitle{#1}}
\def\Lauthor #1{\gdef\LLauthor{#1}}
\def\Laddress #1{\gdef\LLaddress{#1}}
\long\def\Labstract #1{\gdef\LLabstract{#1}}
\gdef\LLabstract{}
\gdef\Lnote #1{\footnote{#1}}

\def\Lmaketitle{
 \thispagestyle{empty}
 \null
 \begin{flushright}
 \LLheader \\
 \LLdate
 \par
 \end{flushright}
 \vskip 4em
 \begin{center}
 {\LARGE\bf \LLtitle \par}
 \vskip 4em
{\large\bf \LLauthor \\}
 \vskip 1em
 {\em \LLaddress}
 \par
 \vskip 4em
 {\bf Abstract} \\
 \end{center}
 \vskip -\the\parskip
 \LLabstract
 \newpage
 \pagestyle{plain}
 \pagenumbering{arabic}
 \gdef\LLheader{}\gdef\LLtitle{}\gdef\LLauthor{}
 \gdef\LLaddress{}\gdef\LLabstract{}
 \let\Lnote\relax
 \let\Lmaketitle\relax}

\def\LLfnsymbol#1{\ifcase#1\or \star\or *\or \dagger\or \ddagger\or
   \mathchar "278\or \mathchar "27B\or \|\or **\or \dagger\dagger
   \or \ddagger\ddagger \else\@ctrerr\fi\relax}

\newfont{\tinyroman}{cmr5 scaled 900}
\newfont{\tinymathsym}{cmsy5 scaled 900}
\newcommand{\tp}{{\mbox{\tinyroman \char'053}}}
\newcommand{\tm}{{\mbox{\tinymathsym \char'000}}}
\newcommand{\tpm}{{\mbox{\tinymathsym \char'006}}}

\newfont{\tinymath}{cmmi10 scaled 600}
\newfont{\verytinymath}{cmmi10 scaled 400}
\newcommand{\tU}{ {\mbox{\tinymath U}} }
\newcommand{\ttU}{ {\mbox{\verytinymath U}} }

\newcommand{\real}{\mbox{\boldmath $R$}}
\newcommand{\integer}{\mbox{\boldmath $Z$}}
\newcommand{\complex}{\mbox{\boldmath $C$}}

\newcommand{\id}{\mbox{$1\hspace{-0.9
ex}1$\hspace{-0.85ex}\raisebox{1.25ex}{\tiny{-}}}\,}

\newfont{\twelveeufm}{eufm10 scaled\magstep1}
\newcommand{\dom}{ \,\mbox{\twelveeufm D} }
\newcommand{\hil}{ \mbox{$\cal H$}\: }
\newcommand{\dd}{\partial}
\newcommand{\rchi}{\raisebox{.4ex}{$\chi$}}
\newcommand{\gal}{\gamma_{\mbox{\tiny L}}}
\newcommand{\gak}{\gamma_{\mbox{\tiny K}}}

\begin{document}


\Lheader{DAMTP R95/1}
\Ldate{January 1995, Revised June 1995}
\Ltitle{Initial Value Problems and\\ Signature Change}
\Lauthor{L. J. Alty$^{a}$\Lnote{\phantom{i}e-mail address:
lja13@amtp.cam.ac.uk}
and C. J. Fewster$^{a,b}$\Lnote{\phantom{i}e-mail address:
fewster@butp.unibe.ch}}
\Laddress{$^{a)}$Department of Applied Mathematics and Theoretical Physics,\\
University of Cambridge, Cambridge CB3 9EW, United Kingdom. \\
{$^{b)}$Institut f\"{u}r Theoretische
Physik, Universit\"{a}t Bern, \\
Sidlerstrasse 5, CH-3012 Bern, Switzerland}}
\Labstract{We make a rigorous study of classical field equations on a
2-dimensional signature changing spacetime using the techniques of operator
theory. Boundary conditions at the surface of signature change are determined
by forming self-adjoint extensions of the Schr\"odinger Hamiltonian. We show
that the initial value problem for the Klein--Gordon equation on this
spacetime is ill-posed in the sense that its solutions are unstable.
Furthermore, if the initial data is smooth and compactly supported away from
the surface of signature change, the solution has divergent $L^2$-norm after
finite time.}

\newpage
\phantom{a}
\vspace{-3em}
\begin{flushright}
gr-qc/9501026
\end{flushright}
\vspace{-4.25em}
\Lmaketitle


\section{Introduction}
\setcounter{equation}{0}

A natural generalisation of general relativity is obtained by relaxing the
restriction that the metric tensor is everywhere Lorentzian. This subject has
produced a spirited debate on the nature of the junction conditions which must
be satisfied by the metric tensor and matter fields at the surface of
signature change
\cite{GibbHart,TGRG,Ellis,Hayward,Hay2,Tucker,KK,KK2,KK3,Alty}.   Part of the
difficulty in finding a consensus in this debate is that much depends on what
one regards as pathological.

One approach \cite{Hayward,Hay2,KK,KK2,KK3} has been to assume certain
differentiability conditions on the spacetime metric and matter fields and
then choose junction conditions which provide a bounded energy momentum
tensor. For example, consider a discontinuously signature changing metric that
satisfies the Einstein equations, and assume that the metric is $C^2$
everywhere (except at the surface of signature change where only the induced
metric is required to be $C^2$). Then the energy momentum tensor is bounded at
the surface of signature change if and only if the second fundamental form of
that surface vanishes\footnote{For a continuously signature changing metric,
the  conditions are slightly stronger, and we recommend that the reader notes
the rigorous work in \cite{KK2} (in particular see theorems~3 and~4).}
\cite{GibbHart,Hayward,KK2}.  As a second example, consider a scalar field
propagating on a spacetime that continuously changes signature. Assuming that
the scalar field is everywhere $C^3$ then the component of its momentum normal
to the surface of signature change must vanish at that surface (see lemma 3
in~\cite{KK3}). This also implies that the energy momentum tensor of the scalar
field is bounded at the surface of signature change.

A second approach \cite{TGRG,Tucker,Alty} (and the one we will adopt in this
paper) is to relax some of the assumptions under which the above junction
conditions are derived. We will study signature change, with no
coupling between gravity and matter, under the assumption that the matter
fields are continuous, but not necessarily differentiable, at the surface of
signature change\footnote{This assumption is not unreasonable and there are
many parallels in physics; for example, a sound wave that passes between two
different media is only $C^0$ at the interface between the media.}.  In this
situation, there are no pre-ordained junction conditions on the
matter fields at the surface of signature change.   This philosophy was used
in~\cite{Alty}, where a variety of different junction conditions were used to
study the propagation of classical fields on a fixed spacetime whose signature
changed from Lorentzian to Kleinian. It was shown that a wide range of field
behaviour could be generated in this way.

Ultimately, of course, one wants to choose the junction conditions which best
model the physical scenario. Unfortunately, the scattering properties of a
surface of signature change are unknown, and in this situation, it makes sense
to choose boundary conditions which provide the most well-behaved system. If,
with such a choice, the system is still poorly defined then one has grounds
for concluding that the situation which is being modelled is unphysical.
However, if the system is fairly well-defined, one should
obtain some reasonable physical predictions, which can then be used to
reassess the theory.  One method of obtaining a well-behaved system is to
impose smoothness conditions as mentioned above. In this paper, we investigate
another possibility which utilises Hilbert space techniques.

For simplicity, we study a 2-dimensional spacetime with signature change from
$(+-)$ to $(++)$, and examine classical fields that propagate, with no
interaction, on the background spacetime\footnote{The background spacetime
that we study satisfies the vacuum Einstein equations exactly (ie.\ there are
no distributional terms at the surface of signature change). This can be
checked either by direct calculation or by using theorem~4 of \cite{KK2}.}. We
identify $+$ with time, and therefore this type of signature change has more
similarity with a Lorentzian to Kleinian signature
transition~\cite{Alty} than a Lorentzian to Riemannian signature transition.
The existence of a global time parameter allows us to write the matter field
equations in Hamiltonian form. We then define the matter field Hamiltonians as
Hilbert space operators so that any junction conditions which are applied to a
matter field are equivalent to {\it boundary conditions} satisfied by
functions in the domain of its Hamiltonian.

In order to obtain a well behaved system, we require the Schr\"{o}dinger
Hamiltonian (ie.\ minus the Laplacian on the background spacetime) to be
self-adjoint, and we also require the boundary conditions to take a general
form which enforces continuity of the functions in its domain. This choice
proves to be highly successful, as it singles out a unique set of boundary
conditions for the matter fields. These boundary conditions were previously
isolated in~\cite{Tucker} using rather different methods. Moreover,
self-adjointness allows us to construct a complete set of (generalised)
eigenfunctions for the Laplacian, which facilitates the construction of
$L^2$-solutions to the Schr\"{o}dinger and Klein--Gordon equations.

In section~2 we briefly review the necessary operator theory, before
proceeding in section~3 to examine the properties of the Dirac, Schr\"odinger
and Klein--Gordon Hamiltonians on the signature changing spacetime. Using
these properties, we explain why our methods cannot be applied to the Dirac
Hamiltonian. In section~4 we determine the boundary conditions corresponding
to self-adjoint extensions of the Schr\"odinger Hamiltonian. Solutions of the
Schr\"odinger and Klein--Gordon equations are constructed and analysed in
section~5. We show that the initial value problem for the Klein--Gordon
equation is ill-posed in the sense that its solutions are unstable, and that
the evolution of reasonable initial data leads to runaway solutions.  In
section~6 we generalise our results and show that they hold for a large class
of signature changing metrics.

\section{Operator Theory}
\setcounter{equation}{0}

Physical systems are often modelled using operators on a Hilbert space.
If such
a system exhibits singular points (where the formal expression of an operator
does not apply in an obvious way), it is often initially convenient to define
the action  of the operators on a space of `nice' functions supported away
from
the singular points.  In our case of interest, this occurs at the
surface of signature change; although it is clear how to define our
operators away from this surface, it is not {\em a priori} clear how to define
them on the surface. The initial choice of domain is not generally sufficient
to effectively model the physical situation, in particular, the operator may
not be self-adjoint when restricted to this domain. If self-adjointness is
required, one can try to extend the domain and action of our operator in order
to determine a rigorously defined self-adjoint operator, which is called a
{\em self-adjoint extension} of the original operator. For differential
operators, this is usually equivalent to the specification of appropriate
boundary conditions at singular points (in our case, at the surface of
signature change).

\subsection{Preliminary definitions}

Although they are standard (see eg.\ \cite{R&SII}),
the following definitions have been included so that the work presented
here is self-contained.

Let ${A}$ be an operator defined on a dense subspace ${\cal D}$ of a
Hilbert space $\hil$ with an inner product $\langle \cdot | \cdot \rangle$
which is anti-linear in the first slot and linear in the second slot.

{\it Definition.}
If for some $\phi\in\hil$
there exists $\eta$ such that
$\langle \phi | {A} \psi \rangle = \langle \eta | \psi \rangle$ for all
$\psi \in {\cal D}$, then $\phi$ is said to be in the domain of the
{\it adjoint} ${A}^*$ of ${A}$, and ${A}^*$ is defined on $\phi$ by
${A}^*\phi = \eta$.

{\it Definition.}
Let ${A}^\prime$ be any operator whose domain includes ${\cal D}$,
then ${A}^\prime$
will be called an
{\it extension} of ${A}$ if
${A}^\prime\psi={A}\psi$ for all $\psi\in {\cal D}$.
${A}$ is said to be {\it symmetric}
if ${A}^*$ is an extension of
${A}$, (ie.\ if
$\langle{\psi}|{{A}\phi}\rangle=\langle{{A}\psi}|{\phi}\rangle$
for all $\psi,\phi\in {\cal D}$).

{\it Definition.}
${A}$ is said to be {\it self-adjoint} if
${A}={A}^*$,
(ie.\ if ${A}$ is symmetric and has the same domain as its adjoint).

{\it Definition.}
If the domain of ${A}^*$ is dense, then we define the
{\it closure} $\bar{{A}}$ of ${A}$ by
$\bar{{A}} = {A}^{**}$.
${A}$ is said to be {\it essentially self-adjoint} if
$\bar{{A}}$
is self-adjoint. Moreover if ${A}$ is essentially self-adjoint
then $\bar{{A}} = {A}^*$, and ${A}^*$ is
the unique self-adjoint extension of ${A}$.

Let us emphasise that an operator (on a Hilbert space) does not exist
independently of either its domain or the Hilbert space on which it is
defined.  For example, when one says that the operator $A$ is self-adjoint, it
is implicit that one is working with the the domain ${\cal D}$ and the Hilbert
space $\hil$ on which $A$ is defined, as well as the inner product
$\langle \cdot | \cdot \rangle$ of $\hil$.

\subsection{Deficiency indices}

The von Neumann theory of deficiency indices \cite{R&SII} can be used to
determine whether a symmetric operator ${A}$ with domain ${\cal D}$
has any self-adjoint extensions, and if so, to construct them.

We define the {\it deficiency subspaces} ${\cal K}^\tp$ and ${\cal K}^\tm$ by
\begin{equation}
{\cal K}^\tpm = \ker({A}^*\mp i ) \; ,
\end{equation}
(ie.\ ${\cal K}^\tpm$ is the subspace of
$L^2$-solutions of equation ${A}^*\Psi = \pm i\Psi$) and define the
{\it deficiency indices} by $n^\tpm = \dim({\cal K}^\tpm)$. There
are three possible cases:

\begin{tabular}{lrcl}
\hspace*{1ex}(1) & \hspace*{1ex}$n^\tp \neq n^\tm$ & $\Longleftrightarrow$ &
$A$ has no self-adjoint extensions. \\
\hspace*{1ex}(2) & \hspace*{1ex}$n^\tp = n^\tm = 0^{\phantom{\tm}}$
& $\Longleftrightarrow$ & $A$ has a unique self-adjoint extension. \\
\hspace*{1ex}(3) & \hspace*{1ex}$n^\tp = n^\tm = N > 0^{\phantom{\tm}}$
& $\Longleftrightarrow$ & $\exists$ a unique family of self-adjoint \\
&&& extensions of $A$ labelled by $U(N)$.
\end{tabular}

In case (2), ${A}$ is essentially self-adjoint and its unique
self-adjoint extension is equal to ${A}^*$.
In case (3), which will be our case of interest,
the self-adjoint extensions are parametrised as follows:
representing $U(N)$ as
the space of unitary maps from ${\cal K}^\tp$ to ${\cal K}^\tm$,
we define for each $U\in U(N)$
the domain ${\cal D}_\tU$ by
\begin{equation}
{\cal D}_\tU = \{ \psi + \rchi^\tp + U\rchi^\tp
| \psi \in {\cal D}, \rchi^\tp \in {\cal K}^\tp \} \; .
\end{equation}
The operator ${A}^*|_{{\cal D}_\ttU}$ is clearly an extension of
${A}$, moreover, ${A}^*|_{{\cal D}_\ttU}$
is essentially self-adjoint with unique
self-adjoint extension ${A}_\tU$ given by
\begin{equation}
{A}_\tU = \left({A}^*|_{{\cal D}_\ttU}\right)^* \; .
\end{equation}
The full set of self-adjoint extensions of ${A}$ is then provided
by the family $\{ {A}_\tU \: |\: U\in U(N) \}$.

\section{Hamiltonian Form of the Classical Wave Equations}
\setcounter{equation}{0}

We will consider a simple 2-dimensional model in which the spacetime
signature changes from $(+-)$ to $(++)$.  We identify `$+$' with time; thus
this type of signature change, in which a spatial coordinate turns into a time
coordinate, has more similarity with four-dimensional Lorentzian to Kleinian
signature change \cite{Alty}, than with four-dimensional Lorentzian to
Riemannian signature change.

{}To begin our investigation we will examine discontinuous signature change
with the flat metric
\begin{equation}
ds^2 = g_{ab} dx^a dx^b = dt^2 + {\rm sign}(z) dz^2 \; .
\label{dcmetric}
\end{equation}
We show in section~6 that the results gained by studying
this metric also apply to more general signature change metrics, including
metrics which exhibit continuous signature change.

Let us briefly consider the measure density appropriate to surfaces of constant
$t$ in this and similar metrics. Since our intention is to use Hilbert space
techniques, we seek a real, positive measure. If $g_{11}$ were
everywhere negative (ie.\ the usual Lorentzian case), one would use
$(-g_{11})^{1/2}dz$. However, for signature changing metrics, this measure
becomes complex, and hence we employ the measure $|g_{11}|^{1/2}dz$, which
is real and positive. For the metric~(\ref{dcmetric}), this reduces to $dz$.
It is suggested in~\cite{Tucker} that one could allow the orientation of the
volume element to change across the surface of signature change, which
would lead to a measure ${\rm sign}(z)|g_{11}|^{1/2}dz$. The inner product
space defined using this measure would be a {\em Krein space}~\cite{Bognar}
rather than a Hilbert space. Analysis in such indefinite inner product spaces
is considerably more involved than in Hilbert spaces, and we will not discuss
this possibility further in this paper.

\subsection{The Dirac Hamiltonian}

In ordinary Minkowski space one can easily define a self-adjoint Dirac
Hamiltonian \cite{Thaller}. We now show that this is not possible
in the flat signature changing spacetime (\ref{dcmetric}). In this section,
the Hilbert space of spinors is taken to be $L^2(\real,dz)\otimes\complex^4$
with its usual inner product. Where no confusion can arise, we will often
write $L^2(\real)$ for $L^2(\real,dz)$. In addition, for any subset $S$ of
$\real$, we denote the space of smooth functions compactly supported in $S$
by $C_0^\infty(S)$.

For $z<0$ define the gamma matrices $\gal^a$ by
$\{\gal^a,\gal^b\} = 2g^{ab}$, and similarly, for $z>0$ define the
Kleinian gamma matrices
$\gak^a$ by $\{\gak^a,\gak^b\} = 2g^{ab}$. Choosing $\gak^0 = \gal^0$,
there are two possibilities for $\gak^1$,
given by $\gak^1 = \pm i\gal^1$.  We will use the standard representation,
in which
\begin{equation}
\gal^0 = \left( \begin{array}{rr} 1 & 0 \\ 0 & -1 \end{array} \right)
\;\;\;\;\;{\rm and}\;\;\;\;\;
\gal^1 = \left( \begin{array}{rr} 0 & 1 \\ -1 & 0 \end{array} \right)\; .
\end{equation}

The Dirac equation for the signature changing
spacetime is
\begin{eqnarray}
(\gal^a\dd_a + im)\Psi = 0 & \phantom{888888} & z<0 \; , \\
(\gak^a\dd_a + im)\Psi = 0 & \phantom{888888} & z>0 \; .
\end{eqnarray}
Defining
\begin{equation}
\xi_\pm(z) =
\left\{
\begin{array}{rcr}
1 & \phantom{888888} & z<0 \; \phantom{,} \\
\pm i & \phantom{888888} & z>0 \; ,
\end{array}
\right.
\end{equation}
the Dirac equation may be written in the Hamiltonian form
\begin{equation}
i \frac{\dd}{\dd t} \Psi = D_\pm \Psi \; ,
\end{equation}
where the two candidate Dirac Hamiltonians
\begin{equation}
D_\pm = \gal^0 \left[ m - i \xi_\pm(z) \gal^1 \frac{\dd}{\dd z} \right]
\end{equation}
correspond to the two possible choices of Kleinian gamma matrices.

Initially, we define $D_\pm$ on the dense domain ${\cal D}
=C_0^\infty(\real\backslash\{0\})\otimes\complex^4$ of smooth spinors
compactly supported away from the origin in the Hilbert space
$L^2(\real)\otimes\complex^4$. A short integration by parts argument
shows that
\begin{equation}
D_\pm^*|_{{\cal D}} =
\gal^0 \left[ m - i \xi_\pm(z)^* \gal^1 \frac{\dd}{\dd z} \right] = D_\mp\; .
\end{equation}
Thus neither $D_+$ nor $D_-$ is symmetric; accordingly, neither admits any
self-adjoint extensions.  Since any reasonable local Dirac Hamiltonian should
certainly contain ${\cal D}$ in its domain, we conclude that the spacetime
(\ref{dcmetric}) does not admit a self-adjoint Dirac Hamiltonian.

\subsection{The Schr\"odinger Hamiltonian}

The two candidate Dirac Hamiltonians both square to give the same operator,
namely
\begin{equation}
D_\pm^2 = {(H+m^2)} \otimes \id
\end{equation}
on $C^\infty_0 (\real\backslash \{0\}) \otimes\complex^4$ in
$L^2(\real)\otimes\complex^4$, where the operator $H$ is given by
\begin{equation}
{H} = {\rm sign}(z) \frac{\dd^2}{\dd z^2}
\label{Schr}
\end{equation}
with domain $C^\infty_0 (\real\backslash \{0\}) \subset L^2(\real)$.
Note that since we are
only concerned with functions that are compactly supported away from the
origin, we need not concern ourselves with derivatives of $\xi(z)$ at $z=0$.
We will refer to $H$ as the Schr\"odinger Hamiltonian.
Since $-\frac{\dd^2}{\dd z^2}$ is unbounded
from above on $L^2(\real^\tm)$ and $\frac{\dd^2}{\dd z^2}$
is unbounded from below on $L^2(\real^\tp)$, we note that
$H$ is unbounded from both above and below.

The fact that ${H}$ is symmetric can be demonstrated
with an integration by parts argument, and hence we can construct the
self-adjoint extensions of ${H}$ using the von Neumann theory
of deficiency indices.

\subsection{The Klein--Gordon Hamiltonian}

The Klein--Gordon equation on the discontinuously
signature changing spacetime (\ref{dcmetric}) is

\begin{equation}
\left(g^{ab} \frac{\dd}{\dd x^a} \frac{\dd}{\dd x^b} + m^2\right) \phi = 0\; .
\label{2ndKG}
\end{equation}

Let
\begin{equation}
\Phi(t) =
\left(
\begin{array}{c} \phi \\ \dot{\phi} \end{array}
\right) \; ,
\end{equation}
where $\dot{\phi} = \frac{\dd\phi}{\dd t}$, then we may write the
Klein--Gordon equation in `first-order form' as
\begin{equation}
\frac{\dd \Phi}{\dd t} = {K} \Phi \; ,
\label{1stKG}
\end{equation}
where the operator
\begin{equation}
{K} =
\left(
\begin{array}{cc} 0 & 1 \\ -{(H+m^2)} & 0 \end{array}
\right)
\end{equation}
on $C^\infty_0 (\real\backslash \{0\}) \oplus L^2(\real) \subset
L^2(\real) \oplus L^2(\real)$ will be called the Klein--Gordon Hamiltonian.

Clearly ${K}$ is not symmetric (on this domain and Hilbert space)
and therefore does not admit any self-adjoint extensions. Despite this,
the special form of $K$ and the fact that $H$ has self-adjoint
extensions will allow us to construct rigorously defined solutions
to~(\ref{1stKG}).

\section{Boundary Conditions}
\setcounter{equation}{0}

In this section we construct self-adjoint extensions of the Schr\"odinger
Hamiltonian ${H}$, and study the boundary conditions which
provide these extensions.  There are many possible sets of boundary conditions
for a matter field $\phi$,
however, we are interested in the subset which describe
how $\phi$ and $\frac{\dd \phi}{\dd z}$ behave as they cross
the surface of signature change.
Since we have no data to guide us, a
reasonable physical assumption is that the field $\phi$ is everywhere
continuous.
This leaves the behaviour of $\frac{\dd \phi}{\dd z}$ to be determined.
Thus we are looking for boundary conditions of the form
\begin{eqnarray}
\phi (0^\tm) &=& \phi (0^\tp) \; , \nonumber \\
&& \label{niceBC} \\
\frac{\dd \phi}{\dd z}(0^\tm) &=& \omega\: \frac{\dd \phi}{\dd z}(0^\tp)
\nonumber
\end{eqnarray}
for some $\omega\in\complex$, $\omega\neq 0$.

\subsection{Self-adjoint extensions}

We will now construct the self-adjoint extensions of ${H}$.
Applying the von Neumann theory of deficiency indices we try to find
$L^2$-bases for ${\cal K}^\tp$ and ${\cal K}^\tm$. We find that ${\cal K}^\tp$
is spanned by
\begin{equation}
\rchi^\tp_1 = \left\{ \begin{array}{lr}
2^{\frac{1}{4}} \exp(-{\rm e}^{i3\pi / 4}z) & z<0 \\
0 & z>0
\end{array} \right.
\;\;
\rchi^\tp_2 = \left\{
\begin{array}{lr}
0 & z<0 \\
2^{\frac{1}{4}} \exp(-{\rm e}^{i\pi / 4}z) & z>0
\end{array} \right.
\label{basis1}
\end{equation}
and ${\cal K}^\tm$ is spanned by
\begin{equation}\label{basis2}
\rchi^\tm_1 = \left\{
\begin{array}{lr}
2^{\frac{1}{4}} \exp(+{\rm e}^{i\pi / 4}z)^{\phantom{3}} & z<0 \\
0 & z>0
\end{array} \right.
\;\;
\rchi^\tm_2 = \left\{
\begin{array}{lr}
0 & z<0 \\
2^{\frac{1}{4}} \exp(+{\rm e}^{i3\pi / 4}z) & z>0
\end{array} \right.
\end{equation}
where the normalisation
$||\rchi^\tp_1|| = ||\rchi^\tp_2|| = ||\rchi^\tm_1|| = ||\rchi^\tm_2|| = 1$
has been implemented.
The deficiency indices are thus $n^\tp = n^\tm = 2$, and hence there is a
family of self-adjoint extensions of ${H}$ labelled by $U(2)$.

As described in section~2, for each $U\in U(2)$ we have a unique
self-adjoint extension ${H}_\tU$ of ${H}$ given by
\begin{equation}
{H}_\tU = \left({H}^*|_{{\cal D}_\ttU}\right)^*
\end{equation}
where
\begin{equation}
{\cal D}_\tU = \{ \phi + \rchi^\tp + U\rchi^\tp
| \phi \in C^\infty_0(\real\backslash \{0\}),
\rchi^\tp \in {\cal K}^\tp \} \; .
\end{equation}

We now construct the boundary conditions which correspond to
${H}_\tU$. For any $\rho,\psi\in\dom({H}^*)$
it is possible to show that
\begin{equation}
\langle \rho | {H}^* \psi \rangle =
\langle {H}^* \rho | \psi \rangle
+ \lim_{\;\; z\rightarrow 0^\tm}
\left[ \frac{\dd \rho^\dagger}{\dd z} \psi -
\rho^\dagger\frac{\dd\psi}{\dd z} \right]
+ \lim_{\;\; z\rightarrow 0^\tp}
\left[ \frac{\dd \rho^\dagger}{\dd z} \psi -
\rho^\dagger\frac{\dd\psi}{\dd z} \right] \; .
\label{terms}
\end{equation}
We note in passing that if we were studying $-\frac{\dd^2}{\dd z^2}$ instead
of $H={\rm sign}(z)\frac{\dd^2}{\dd z^2}$ on
$C^\infty_0(\real\backslash\{0\})\subset L^2(\real)$, then the boundary terms
in the analogue to (\ref{terms}) would differ by a relative sign \cite{Seba}.

The domain of ${H}_\tU$ consists of those $\rho\in\dom (H^*)$ for which the
boundary terms in (\ref{terms}) vanish for all $\psi\in{\cal D}_\tU$.
Clearly for $\psi\in C^\infty_0(\real\backslash\{0\})$ the
boundary terms are zero, hence it suffices to consider
$\psi\in \{ \rchi^\tp + U\rchi^\tp | \rchi^\tp \in {\cal K}^\tp \}$. By
choosing such $\psi$ we are able to determine the boundary conditions which
$\rho^\dagger$ and $\frac{\dd \rho^\dagger}{\dd z}$ must satisfy at $z=0$.

Choosing the basis (\ref{basis1}) for ${\cal K}^\tp$ and the
basis (\ref{basis2}) for ${\cal K}^\tm$, the unitary map $U$ can be
written as a matrix
\begin{equation}
U = \left( \begin{array}{cc} a&b\\ c&d \end{array} \right)
\label{eq:U}
\end{equation}
where
\begin{equation}
a\bar{a} + b\bar{b} = 1 \; , \;\;\;
c\bar{c} + d\bar{d} = 1 \; , \;\;\;
a\bar{c} + b\bar{d} = 0 \; .
\label{conditions}
\end{equation}
By substituting for $\psi$ in (\ref{terms}), it can now be shown that
the boundary terms vanish if and only if
\begin{equation}
A
\left( \begin{array}{c} \frac{\dd \rho^\dagger}{\dd z}(0^\tm) \\
 \\
\frac{\dd \rho^\dagger}{\dd z}(0^\tp) \end{array} \right)
= {\rm e}^{i\pi /4} B
\left( \begin{array}{c} \rho^\dagger (0^\tm) \\
 \\
\rho^\dagger (0^\tp)  \end{array} \right)
\label{generalBC}
\end{equation}
where the matrices $A$ and $B$ are given by
\begin{equation}
A =
\left( \begin{array}{cc}
a+1 & c \\
 & \\
b & d+1
\end{array} \right)
\;\;\;\; {\rm and} \;\;\;\;
B =
\left( \begin{array}{cc}
a-i & ic \\
 & \\
b & id-1
\end{array} \right) \; .
\label{AB}
\end{equation}
Thus for each unitary matrix $U$ there is a self-adjoint extension
${H}_\tU$ of ${H}$ with a corresponding set of boundary conditions
(\ref{generalBC}).

Finally, we note that the
addition of a constant term to ${H}$ does not alter this class of
boundary conditions, and therefore the
self-adjoint extensions of $H+m^2$ follow immediately from those of ${H}$.

\subsection{Determining the boundary conditions}

We wish to find self-adjoint extensions of ${H}$ with corresponding boundary
conditions of the form (\ref{niceBC}).  In order to complete this task, it is
clear from (\ref{generalBC}) that $A$ and $B$ must be singular, since if either
$A$ or $B$ is invertible, we will obtain boundary conditions which relate the
field to its derivative.  In the appendix we prove the surprising result that
there is only one self-adjoint extension satisfying the above restrictions. Its
corresponding boundary conditions are
\begin{eqnarray}
\rho (0^\tm) &=& \rho
(0^\tp) \; , \nonumber\\ && \label{finalBCs} \\
\frac{\dd \rho}{\dd z}(0^\tm)
&=& -\:\frac{\dd \rho}{\dd z}(0^\tp) \; .  \nonumber
\end{eqnarray}
This is the set of boundary conditions which we will adopt in section~5 for
the construction of wave equation solutions. These boundary conditions were
previously isolated, using rather different methods, in~\cite{Tucker}. In fact,
the authors of~\cite{Tucker} also consider the case where the orientation of
the volume form is different on the two sides of the boundary, in which case
the relative minus between the normal derivatives $\frac{\dd\rho}{\dd z}
(0^\tpm)$ is removed.  As we mentioned in section~3, we exclude this
possibility from our treatment as we require the integration measure to be
both real and positive.

In \cite{Alty} particular attention was devoted to the boundary conditions
$\rho^\dagger$ continuous at $z=0$,
and either $\frac{\dd \rho}{\dd z}(0^\tm)
= \frac{\dd \rho}{\dd z}(0^\tp)$
or $\frac{\dd \rho}{\dd z}(0^\tm)
= \pm i\frac{\dd \rho}{\dd z}(0^\tp)$; it is worth emphasising that
no self-adjoint extension corresponds to such boundary conditions.

\section{Wave Equation Solutions}
\setcounter{equation}{0}

In this section we construct solutions to the Schr\"odinger
and Klein--Gordon equations on the signature changing spacetime
(\ref{dcmetric}),
satisfying the boundary conditions
(\ref{finalBCs}). Throughout this section ${H}_\tU$ will
represent the self-adjoint extension of ${H}$ corresponding
to these boundary conditions, and $\hil$ will denote the Hilbert space
$L^2(\real,dz)$.

\subsection{Solutions of the Schr\"odinger equation}

For our choice of self-adjoint extension, the Schr\"odinger equation on the
signature changing spacetime~(\ref{dcmetric}) is
\begin{equation}
i\frac{\dd \Psi}{\dd t} = {H}_\tU \Psi \; ,
\label{SchrEq}
\end{equation}
and has general solution $\Psi(t) = {\rm e}^{-iH_{\ttU}t}\Psi(0)$. The
evolution operator $e^{-iH_{\ttU}t}$ is most conveniently constructed using
the
spectral representation of $H_\tU$, which we now study.

Solving the time-independent Schr\"{o}dinger equation
\begin{equation}
{H}_\tU \tilde{\Psi} = E \tilde{\Psi}
\label{TransSchrEq}
\end{equation}
as an ODE subject to the boundary conditions (\ref{finalBCs}), we obtain the
generalised eigenfunctions (mode solutions)
\begin{eqnarray}
\tilde{\Psi}(z,E) = &a(E)& \!\!\!\left\{
(1-i)\:\theta(-z)\theta(-E)\:\exp(\sqrt{-E}z) \right. \nonumber \\
&+&(1-i)\:\theta(+z)\theta(-E)\left[\cos(\sqrt{-E}z) - \sin(\sqrt{-E}z)\right]
\nonumber \\
&+&(1+i)\:\theta(-z)\theta(+E)\left[\cos(\sqrt{+E}z) + \sin(\sqrt{+E}z)\right]
\nonumber \\
&+& \left. (1+i)\:\theta(+z)\theta(+E)\:\exp(-\sqrt{+E}z) \right\} \; ,
\end{eqnarray}
where $\theta$ is the usual step function and the function a(E) is
\begin{equation}
a(E) =  \frac{1}{\sqrt{2\sqrt{|E|}}} \; .
\end{equation}

Next, we define an integral transform ${\cal M}:L^2(\real,dE)\rightarrow
L^2(\real,dz)$ by
\begin{equation}
({\cal M} f)(z) = \frac{1}{\sqrt{2\pi}} \int^\infty_{-\infty}
f(E) \tilde{\Psi}(z,E) dE \; .
\end{equation}
In appendix B, we show that ${\cal M}$ is unitary. Heuristically, this
is equivalent to the continuum normalisation relations
\begin{equation}
\int_{-\infty}^\infty dz\; \tilde{\Psi}(z,E)^*\tilde{\Psi}(z,E^\prime)
= 2\pi\delta(E-E^\prime)\; ,
\end{equation}
and
\begin{equation}
\int_{-\infty}^\infty dE\; \tilde{\Psi}(z,E)^*\tilde{\Psi}(z^\prime,E)
= 2\pi\delta(z-z^\prime)\; .
\end{equation}

A simple calculation shows that ${\cal M} E {\cal M}^{-1}$ agrees with $H$ on
$C_0^\infty(0,\infty)$. It is therefore one of the self-adjoint extensions
of $H$, and furthermore it is clear that it must be equal to $H_\tU$. Thus
we have
\begin{equation}
H_\tU = {\cal M} E {\cal M}^{-1}\; ,
\end{equation}
and using the spectral
theorem~\cite{R&SI}, we see that the general solution of (\ref{SchrEq})
satisfying the boundary conditions (\ref{finalBCs}) may be written in
initial value form as
\begin{equation}
\Psi(z,t) = {\cal M} {\rm e}^{-iEt} {\cal M}^{-1} \Psi(z,0) \; ,
\end{equation}
for any $\Psi(z,0)\in \hil$.  For smooth compactly supported initial data, one
may show that the solutions $\Psi(z,t)$ are $C^{\infty}$ everywhere except at
$z=0$ where they are $C^0$.

It would also be possible to give an explicit form for the integral kernel of
the evolution operator ${\rm e}^{-iH_{\ttU}t}$. However, we will not do this
here because our main focus is the Klein--Gordon equation.

\subsection{Solutions of the Klein--Gordon equation}

Given our choice of Schr\"{o}dinger Hamiltonian $H_\tU$, the Klein--Gordon
Hamiltonian becomes
\begin{equation}
K_{\tU} =
\left(
\begin{array}{cc} 0 & 1 \\ -{(H_{\tU}+m^2)} & 0 \end{array}
\right)
\end{equation}
on $\dom(H_\tU) \oplus L^2(\real) \subset L^2(\real) \oplus L^2(\real)$.
Initially, we consider the massless case ($m=0$) and observe,
following \cite{Bernard}, that the first order form
of the Klein--Gordon
equation
\begin{equation}
\frac{\dd \Phi}{\dd t} = {K_\tU} \Phi
\label{1stKG2}
\end{equation}
has a formal solution of the form
\begin{equation}
\Phi(t) = {\cal T}(t) \Phi(0) \; ,
\label{solution}
\end{equation}
where
\begin{equation}
{\cal T}(t) =
\left(
\begin{array}{cc}
\cos({H}_\tU^{1/2}t) & {H}_\tU^{-1/2}\sin({H}_\tU^{1/2}t) \\
 & \\
-{H}_\tU^{1/2}\sin({H}_\tU^{1/2}t) & \cos({H}_\tU^{1/2}t)
\end{array}
\right) \; . \label{evop}
\end{equation}
At first sight, ${\cal T}(t)$ seems to depend on some choice of square root.
However, as noted in \cite{Bernard}, the formal power series for
$\cos({H}_\tU^{1/2}t)$,
${H}_\tU^{-1/2}\sin({H}_\tU^{1/2}t)$ and
${H}_\tU^{1/2}\sin({H}_\tU^{1/2}t)$
are independent of any definition of the square root. Indeed, we have that
\begin{eqnarray}
\cos({H}_\tU^{1/2}t) = 1 - \frac{{H}_\tU t^2}{2!}
+ \frac{{H}_\tU^2 t^4}{4!} - \cdots \; , \\
{H}_\tU^{-1/2}\sin({H}_\tU^{1/2}t) = t - \frac{{H}_\tU t^3}{3!}
+ \frac{{H}_\tU^2 t^5}{5!} - \cdots \; , \\
{H}_\tU^{1/2}\sin({H}_\tU^{1/2}t)
= {H}_\tU t - \frac{{H}_\tU^2 t^3}{3!}
+ \frac{{H}_\tU^3 t^5}{5!} - \cdots \; .
\end{eqnarray}
Thus ${\cal T}(t)$ can be unambiguously defined using the spectral
theorem, with domain
\begin{equation}
\dom({\cal T}(t)) = \dom({H}_\tU^{1/2}\sin({H}_\tU^{1/2}t)) \oplus
\dom(\cos({H}_\tU^{1/2}t)) \; .
\label{Tdomain}
\end{equation}
Note also that if $\Phi\in\dom({\cal T}(t_1))$ and
${\cal T}(t_1)\Phi\in\dom({\cal T}(t_2))$ then the group property
\begin{equation}
{\cal T}(t_1+t_2) \Phi = {\cal T}(t_2){\cal T}(t_1)\Phi
\end{equation}
is satisfied.

Let $E_{[A,B]}$ denote the spectral projector of ${H}_\tU$ on $[A,B]$
(ie.\ the
space of position space functions with `$A \leq {\rm energy} \leq B$').
By considering the positive and negative energy subspaces of
$H_\tU$, then we see that $\cos({H}_\tU^{1/2}t)$ may be written
\begin{equation}
\cos(({H}_\tU)^{1/2} t) =
\cos(({H}_\tU|_{E_{[0,\infty)}})^{1/2} t) +
\cosh((-{H}_\tU|_{E_{(-\infty,0]}})^{1/2} t) \; .
\label{posneg}
\end{equation}
Similar decompositions clearly exist for
${H}_\tU^{-{1/2}}\sin({H}_\tU^{1/2}t)$ and
${H}_\tU^{1/2}\sin({H}_\tU^{1/2}t)$. Note that it is the exponentially
growing parts of these decompositions which force us to restrict the
domain of ${\cal T}(t)$ to the subset of $\hil\oplus\hil$
given in (\ref{Tdomain}). Thus we see that, in contrast to the bounded
Schr\"odinger evolution ${\rm e}^{-iH_\ttU t}$, ${\cal T}(t)$ is unbounded.

Using (\ref{posneg}) and its analogues, and dominated convergence arguments
similar to those employed in the proof of Theorem VIII.7 in \cite{R&SI}, it is
possible to show that, for any $\tau>0$ and $\Phi(0)\in\dom({\cal T}(\tau))$

\begin{tabular}{ll}
(i) & ${\cal T}(t) \Phi(0) \in \dom(K_{\tU})$ for all $t\in [0,\tau)$,\\
(ii) & The vector-valued function $\Phi(t) = {\cal T}(t)\Phi(0) $ is
differentiable with \\ & respect to $t$ for all $t\in [0,\tau)$,
with derivative $K_{\tU}\Phi(t)$.
\end{tabular}

Thus $\Phi(t)$ is an $L^2$-solution of the Klein--Gordon equation
(\ref{1stKG})  satisfying the boundary conditions (\ref{finalBCs}) and with
initial data $\Phi(0)$.  Note that a different self-adjoint extension would
yield solutions satisfying  different boundary conditions at $z=0$.

\subsection{Initial Data for the Klein--Gordon equation}

{\bf Instability}

It is clear from (\ref{posneg}) and the fact that $H_\tU$ is unbounded from
below that the operator ${\cal T}(t)$ is unbounded.
As a consequence, there exist sequences of
initial data $\Phi_n(0)$ with $\Phi_n(0)\rightarrow 0$ but
${\cal T}(t)\Phi_n(0)$ divergent for any $t>0$.
For example, simply choose any $\{ \Phi_n(0) \}$ such that
$|| \Phi_n(0) || \le \frac{1}{n}$ and
\begin{equation}
\Phi_n(0) \in E_{[-n,1-n]} \hil \oplus E_{[-n,1-n]} \hil \; .
\end{equation}
Clearly $\Phi_n(0)\in \dom({\cal T}(t))$ for all $t\in\real$
and $\Phi_n(0)\rightarrow 0$ as $n\rightarrow\infty$. However,
\begin{equation}
||{\cal T}(t)\Phi_n(0)|| \ge \cosh( (n-1)^{1/2} t) \; ,
\end{equation}
which diverges as $n\rightarrow \infty$. This problem is related to the fact
that initial value problems  for elliptic equations are ill-posed (see eg.\
Hadamard's example in Chapter III \S 6.2 of \cite{C+H}).
However, as is noted in \cite{C+H}, many
problems which are of physical interest are actually ill-posed, and so we will
proceed, but with caution.

{\bf Runaway solutions}

A reasonable experiment would be to observe the
behaviour of matter which propagated through a Lorentzian region and
scattered off a Kleinian region. Such an experiment is modelled by evolving
data which is initially compactly supported in
the Lorentzian region $z<0$.

Consider any smooth initial data $\Phi(0)$ with compact support
$[-z_1,-z_0]$ for some real constants $z_1>z_0>0$. Thus $\Phi(0)$
can be written as
\begin{equation}
\Phi(0)  = \Phi(z,0) = \left\{ \begin{array}{lcl}
\left( \begin{array}{c} \phi_1(z) \\ \phi_2(z) \end{array} \right)
&\;\;\;\;\;\;& z\in[-z_1,-z_0] \\
&& \\
\left( \begin{array}{c} 0 \\ 0 \end{array} \right) &\;\;\;\;\;\;&
{\rm otherwise}
\end{array} \right.
\end{equation}
for smooth compactly supported functions $\phi_1(z)$ and $\phi_2(z)$,
where $z_1=-\inf\{z|\Phi(z,0)\not=0\}$ and $z_0=-\sup\{z|\Phi(z,0)\not=0\}$.

Let us define $\Phi^\tm(E,t)$ to be the negative energy part of the evolved
data at time $t$ in the energy representation, ie.\
\begin{equation}
\Phi^\tm(E,t)= \theta(-E)(({\cal M}^{-1}\oplus {\cal M}^{-1}) \Phi(t))(E) \; ,
\end{equation}
then we find
\begin{equation}
\Phi^\tm(E,t) = k(E)
\left( \begin{array}{c}
\cosh(\sqrt{|E|}t)\hat{\phi}_1(E)
+ \frac{1}{\sqrt{|E|}}\sinh(\sqrt{|E|}t) \hat{\phi}_2(E)  \\
 \\
\sqrt{|E|}\sinh(\sqrt{|E|}t)\hat{\phi}_1(E)
+ \cosh(\sqrt{|E|}t)\hat{\phi}_2(E)
\end{array} \right)\; ,
\label{eq:evo}
\end{equation}
where
\begin{equation}
k(E) = \frac{(1+i)\theta(-E)a(E)}{\sqrt{2\pi}} \; ,
\end{equation}
and the $\hat{\phi}_i(E)$ are defined by
\begin{equation}
\hat{\phi}_i(E) = \int^{-z_0}_{-z_1} \phi_i(z) {\rm e}^{\sqrt{|E|}z} dz
\;\;\;\;\;\;\;\; i=1,2.
\end{equation}

One may easily establish two decay estimates for these functions. Firstly,
for each $N\in\integer$, there exists a constant $C_N$ such that
\begin{equation}
|\hat{\phi}_i(E)| \le \frac{C_N}{1+|E|^{N/2}} e^{-\sqrt{|E|}z_0}
\end{equation}
for all $E<0$. For $N=0$, this is obvious; for general $N$, one
replaces $\phi_i(z)$ by $\phi_i(z)+(-1)^N d^N\phi_i/dz^N$ and applies the
same argument.
Secondly, for all sufficiently small $\epsilon>0$ there exists a
constant $C>0$ such that
\begin{equation}
|\hat{\phi}_i(E)| \ge C e^{-\sqrt{|E|}(z_0+\epsilon)}
\end{equation}
for all sufficiently negative $E$. This estimate is proved by taking
$\epsilon$ small enough so that the real and imaginary parts of
$\phi$ are single-signed on $(-z_0-\epsilon,-z_0)$.

Comparing with~(\ref{eq:evo}), it is clear that $\Phi^\tm(E,t)$ is in
$L^2(\real^\tm,dE)\oplus L^2(\real^\tm,dE)$ for $t\le z_0$, but that it fails
to be $L^2$ for $t>z_0$. Thus the initial data cannot be evolved beyond
the instant at which it begins to arrive at the surface of signature change.
Such behaviour suggests that there is a severe back-reaction just after
$t=z_0$.   Further analysis shows that the solution is
$C^\infty$ for $t<z_0$.

It is easy to see that the re-introduction of mass makes little difference
to our analysis. Essentially, the point is that the addition of a finite mass
term still leaves the Hamiltonian unbounded from below, which is the root of
all the above problems. If we re-define
$\Phi^\tm(E,t)=
\theta(-(E+m^2))(({\cal M}^{-1}\oplus {\cal M}^{-1}) \Phi(t))(E)$, one can see
that this portion of the solution is evolved by factors such as
$\cosh\sqrt{|E+m^2|}t$. The same decay bounds as before show that
$\Phi^\tm(E,t)$ fails to be $L^2$ as soon as $t>z_0$.

\section{Variations and Generalisations}
\setcounter{equation}{0}
\subsection{General Kleinian signature change}

We now consider the class of 2-dimensional metrics (which includes the metric
(\ref{dcmetric})) of the form
\begin{equation}
ds^2 = dt^2 + h(z) dz^2 \; ,
\label{gensc}
\end{equation}
where $h(z)$ is $C^\infty$ and non-zero on all compact
sets which exclude the origin, and satisfies $h(z)={\rm sign}(z)$ outside
a compact neighbourhood of the origin. Thus $h(z)$ is bounded away from the
origin. Initially, we will also assume
that $z|h(z)|^{1/2}\rightarrow 0$ as $z\rightarrow 0$ (as a two-sided limit).
Metrics in this class exhibit precisely one change of signature
from Lorentzian to Kleinian, due to $h(z)$ either
vanishing or becoming (mildly) singular or discontinuous at the origin.

Let $\hil^\prime$ denote the Hilbert space $L^2(\real,|h(z)|^{1/2}dz)$, and
define
\begin{equation}
H^\prime = h(z)^{-1} \left(\frac{\partial^2}{\partial z^2} -
\frac{h^\prime(z)}{2h(z)}\frac{\partial}{\partial z}
\right)
\end{equation}
with domain ${\cal D}^\prime = C_0^\infty(\real\backslash\{0\})\subset
\hil^\prime$. This operator is equal to minus the Laplacian on the
spacetime (\ref{gensc}), and it is therefore the analogue of the operator $H$
on ${\cal D}=C_0^\infty(\real\backslash\{0\})\subset L^2(\real, dz)$ on the
the spacetime (\ref{dcmetric}). We now investigate the self-adjoint
extensions of $H^\prime$.

Due to our assumption that the $z|h(z)|^{1/2}\rightarrow 0$ as a two-sided
limit as $z\rightarrow 0$, we can define a new coordinate $\omega$
by
\begin{equation}
\omega(z) = \int_0^z dz^\prime |h(z^\prime)|^{1/2}
\end{equation}
which has the same sign as $z$ and satisfies
$h(z)dz^2 ={\rm sign}(\omega)d\omega^2$. Thus we have a coordinate
transformation from our new metric to the discontinuous case studied above.
We note that, in general, this transformation is not smooth at $z=0$.

Now consider the unitary operator
$W:L^2(\real, d\omega)\rightarrow \hil^\prime$ implementing the coordinate
change by
\begin{equation}
(W\phi)(z) = \phi(\omega(z)) \; ,
\end{equation}
and note that $W$ preserves the space of smooth functions compactly supported
away from the origin. It is then easy to check that
\begin{equation}
(H^\prime W\phi)(z) = (WH\phi)(z)
\end{equation}
for all $\phi\in C_0^\infty(\real\backslash\{0\})$.
The choice of domain ensures that the smoothness or otherwise of $\omega(z)$
at the origin is irrelevant. We therefore have
$H^\prime = WHW^*$. By the unitarity of $W$,
$H^\prime$ and $H$ have the same deficiency indices and their self-adjoint
extensions are related by
\begin{equation}
H^\prime_\tU= W H_\tU W^* \; ,
\end{equation}
for $U\in U(2)$.

Accordingly, the results of sections~4 and~5 can be translated directly into
results for metrics of the form~(\ref{gensc}) satisfying
$z|h(z)|^{1/2}\rightarrow 0$ as $z\rightarrow 0$ as a two-sided limit. In
particular, because $\dom(H^\prime_\tU) = W\dom(H_\tU)$,  the boundary
conditions corresponding to unitary matrix $U$ are
\begin{equation}
A
\left( \begin{array}{c} |h(z)|^{-1/2}
\frac{\dd \rho^\dagger}{\dd z}(0^\tm) \\
 \\
|h(z)|^{-1/2}
\frac{\dd \rho^\dagger}{\dd z}(0^\tp) \end{array} \right)
= {\rm e}^{i\pi /4} B
\left( \begin{array}{c} \rho^\dagger (0^\tm) \\
 \\
\rho^\dagger (0^\tp)  \end{array} \right) \; ,
\end{equation}
where the matrices $A$ and $B$ are once again given by (\ref{AB}).

The distinguished boundary condition picked out in section~4 therefore
becomes
\begin{eqnarray}
\rho (0^\tm) &=& \rho (0^\tp) \; , \nonumber\\ && \label{distbc} \\
\lim_{z\rightarrow 0^\tm}
|h(z)|^{-1/2}\frac{\dd \rho}{\dd z}(z) &=&
-\:\lim_{z\rightarrow 0^\tp}
|h(z)|^{-1/2}\frac{\dd \rho}{\dd z}(z) \; .
\end{eqnarray}
This boundary condition corresponds to a slight generalisation of one of the
possible choices isolated in~\cite{Tucker}, as it includes the case of
mildly singular $h(z)$.

Turning to section~5, the entire analysis goes through under the unitary
equivalence $W$. The initial value problem is ill-posed for the
distinguished boundary conditions, and the evolved data fails to be $L^2$
immediately after it begins to strike the surface of signature change.

For completeness, let us now relax the condition that $z|h(z)|^{1/2}\rightarrow
0$ as a two-sided limit. The limit may fail from above, below, or both above
and below. The physical interpretation of this is easily understood as follows.
Suppose the limit fails from above. Then any coordinate transformation
$z\mapsto \omega(z)$ with $d\omega = |h(z)|^{1/2}dz$ necessarily maps
$(0,\infty)$ onto the whole real line. This corresponds to the fact that no
causal curve can cross the surface of signature change in finite coordinate
time. Thus this metric is not really a model of signature change, because one
cannot cross the boundary between the two regions. Nonetheless, it is
instructive to see how this alters our analysis.

Writing $H^\prime$ as the direct sum of operators on $L^2(\real^\tpm,
|h(z)|^{1/2}dz)$, we may compute its deficiency indices by considering the
contributions from each half-line separately. Let us consider $\real^\tp$.  If
$z|h(z)|^{1/2}\rightarrow 0$ as $z\rightarrow 0^\tp$, then $H^\prime$ on
$C_0^\infty(0,\infty)$ is unitarily equivalent to
$\frac{\partial^2}{\partial\omega^2}$ on $C_0^\infty(0,\infty)\subset
L^2(\real^\tp,d\omega)$, which has deficiency indices $\langle 1,1\rangle$.
Alternatively, if $z|h(z)|^{1/2}\not\rightarrow 0$ as $z\rightarrow 0^\tp$
then $H^\prime$ on $C_0^\infty(0,\infty)$ is unitarily equivalent to
$\frac{\partial^2}{\partial\omega^2}$ on $C_0^\infty(\real)\subset
L^2(\real,d\omega)$, which is essentially self-adjoint and therefore has
deficiency indices $\langle 0,0\rangle$. Identical arguments for the
negative half-line allow us to build up the full picture as
follows\footnote{Obviously, we could also consider the case in which
$z|h(z)|^{1/2}\rightarrow 0$ from both above and below. This replicates the
results presented at the beginning of this section, ie.\ $H^\prime$ on
$C^\infty_0(\real\backslash 0)$ has deficiency indices $\langle 2,2\rangle$.}:

{\noindent Case} (i): $z|h(z)|^{1/2}\rightarrow 0$ from neither above
nor below. The total deficiency indices of $H^\prime$ are $\langle 0,0\rangle$,
so there is a unique self-adjoint extension, which may be shown to correspond
to Dirichlet boundary conditions at $z=0$.

{\noindent Case} (ii): $z|h(z)|^{1/2}\rightarrow 0$ from below but not
from above. The total deficiency indices of $H^\prime$ are $\langle
1,1\rangle$, so there is a 1-parameter family of self-adjoint extensions
labelled by $L\in\real\cup \{\infty\}$, corresponding to boundary conditions
\begin{equation} \lim_{z\rightarrow 0^\tm} \left(|h(z)|^{-1/2}
\frac{\partial\rho}{\partial z}(z)
 + L^{-1}\rho(z) \right) = 0 \; , \qquad \rho(0^\tp)=0 \; .
\label{1pb}
\end{equation}

{\noindent Case} (iii): $z|h(z)|^{1/2}\rightarrow 0$ from above but not
from below. Again, we have deficiency indices $\langle 1,1\rangle$ and a
1-parameter family of boundary conditions obtained from~(\ref{1pb}) by
reflection about $z=0$.

Cases (i)--(iii) are distinguished by the fact that the two regions $z<0$ and
$z>0$ are decoupled from one another. The surface of signature change is
impenetrable to classical particles and also to classical fields.
The initial value problem is well-posed for smooth initial data compactly
supported in the Lorentzian region (although it is ill-posed for such data
supported in the Kleinian region). However, as we have indicated,
metrics which do not satisfy $z|h(z)|^{1/2}\rightarrow 0$ as $z\rightarrow 0$
are not really examples of signature change, so the well-posedness of this
system is of little interest.

Finally, let us consider the Dirac equation in the spacetime~(\ref{gensc}).
In the notation of section~3, the two candidate Dirac Hamiltonians are
\begin{equation}
D^\prime_\pm = \gal^0 \left[ m - i \xi_\pm(z) \gal^1
|h(z)|^{-1/2} \frac{\dd}{\dd z} \right]
\end{equation}
on ${\cal D}^\prime = C_0^\infty(\real\backslash\{0\})\otimes \complex^4
\subset L^2(\real,|h(z)|^{1/2}dz)\otimes\complex^4$, where the
$\gamma$-matrices $\gal^a$
obey the Minkowski space algebra. As in section~3, we find that
$(D^\prime_\pm)^*|_{{\cal D}^\prime} = D^\prime_\mp$, and therefore neither
operator admits any self-adjoint extensions.

\subsection{Second Order Analysis}

In case it is thought that the runaway solution of section~5 is simply due to
the first order formalism and our use of self-adjoint extensions, we provide a
second order analogue of the result using only Fourier analysis and
elementary arguments. In the following, it will only be necessary to assume
boundary conditions of the general form~(\ref{niceBC}). This allows us to
consider boundary conditions for which both the field and its derivative are
continuous, as well as the boundary conditions employed in section~5.

Consider the Klein--Gordon equation (\ref{2ndKG}) in
the Kleinian region $(z>0)$ with $m=0$. Fourier transforming in $t$ we obtain
\begin{equation} \frac{\dd^2\tilde\phi}{\dd z^2} = E^2 \tilde\phi\; .
\end{equation}
Taking the solution which decays as $z\rightarrow \infty$ we find
\begin{equation}
\tilde\phi (E,z) = \tilde\phi (E,0) {\rm e}^{-|E|z}\; ,
\end{equation}
and hence
\begin{equation}
\frac{\dd\tilde\phi}{\dd z} (E,0^\tp) = -|E| \tilde\phi (E,0) \; .
\end{equation}
Thus, on the Lorentzian side of the signature change surface,
\begin{equation}
\frac{\dd\tilde\phi}{\dd z} (E,0^\tm) = -\omega|E| \tilde\phi (E,0) \; ,
\end{equation}
where $\omega$ parametrises the boundary condition~(\ref{niceBC}).

If $\tilde\phi (E,0)$ is analytic in $E$ (but not identically zero) then we
see that  $\frac{\dd\tilde\phi}{\dd z} (E,0^\tm)$ is not analytic, and
{\it vice
versa}. Furthermore, by the Paley--Wiener theorem (see eg.\ \cite{R&SII}),  we
then have that it is not possible for both $\phi (t,0)$ and
$\frac{\dd\phi}{\dd
z} (t,0^\tm)$ to be elements of $C^\infty_0(\real)$.

Let us again model the experiment described in the section~5. Choose the
initial data $\phi (0,z)$ and $\frac{\dd\phi}{\dd t} (0,z)$ to be in
$C^\infty_0(-\infty,0)$, and assume that at least some of the matter
propagates
towards the Kleinian region at $z=0$.  For massless particles this would give
data $\phi (t,0)$ and $\frac{\dd\phi}{\dd z} (t,0^\tm)$ on the surface of
signature change which are both in $C^\infty_0(\real)$, but such data will not
provide a solution of the Klein--Gordon equation in the Kleinian region which
decays as $z\rightarrow\infty$. Hence the second order formalism also leads to
divergent solutions from smooth compactly supported initial data.

\section{Conclusion}
\setcounter{equation}{0}

We have seen that the Schr\"odinger Hamiltonian on the flat
signature changing spacetime (\ref{dcmetric}) admits
self-adjoint extensions. The additional restrictions that (i) the boundary
conditions corresponding to the self-adjoint extension have the form of
junction conditions for the matter field and its first derivative, and
(ii) that
the matter field is continuous, were shown to pick out one particular set of
boundary conditions (\ref{finalBCs}). Using these results, we were able to
construct solutions of the Schr\"odinger and Klein--Gordon matter
field equations.  The Dirac Hamiltonian is non-symmetric and it is highly
likely that the Dirac equation is only satisfied by solutions of
the type found in \cite{Alty}.

The boundary conditions (\ref{finalBCs}) provide a well-behaved system to
study, and furthermore, since a single set of boundary conditions are
selected, the requirement of self-adjointness provides a possible way of
solving the dilemma of which boundary conditions to choose for signature
change. The boundary conditions (\ref{finalBCs}) are
physically very similar to what might be regarded as the `natural boundary
conditions' where both the field and its derivative are
continuous, since on applying them to mode solutions (as in \cite{Alty}) we
find that a scalar field is completely reflected, whilst the Dirac
field is completely absorbed.

A reasonable physical experiment would be to observe the result of throwing
matter at a Kleinian region. In our model, this type of experiment corresponds
to taking smooth compactly supported initial data in the Lorentzian region
$z<0$ at $t=0$. The results for the Schr\"odinger equation were very promising.
Smooth compactly supported initial data in the Lorentzian region could be
evolved indefinitely.  However, the more physically relevant Klein--Gordon
solutions gave very different results.  Although we could define such
solutions, in section~5 we saw that smooth compactly supported initial data
could only be evolved for a finite time. Indeed, for a massless scalar field
this time was equal to the earliest possible time that classical matter could
reach the surface of signature change. As soon as matter reaches the signature
change surface the spatial integral of the energy momentum tensor of that
matter field becomes unbounded and our assumption that the matter fields do not
interact with the background spacetime breaks down. Moreover, we have seen
that these results apply to general Kleinian signature change
systems.

One possible resolution would be to impose Dirichlet boundary conditions at
the surface of signature change~\cite{Innigo}. These boundary conditions may
also be studied using the self-adjoint extension approach (in fact they
correspond to setting the unitary map (\ref{eq:U}) to $U=-\id$), and they
provide a well-posed system for initial data compactly supported in the
Lorentzian region. The cost of obtaining this well-posed system is that
the regions of differing signature are essentially decoupled. When there is
only one change of signature, this is tantamount to ignoring the Kleinian
region, and it is therefore questionable whether this is a good model for
signature change. However, in systems with more than one change of signature,
it is possible that such boundary conditions might lead to interesting
effects~\cite{Innigo}.

Returning therefore to boundary conditions (\ref{finalBCs}); the
consequent severe behaviour of the Klein--Gordon field suggests
that, if interaction were allowed, the back-reaction of matter fields on the
metric might annihilate the region of signature change, and thus regions of
signature change would be unable to form. These results are consistent with
what could be called the ``Signature Protection Conjecture'', or in other
words, the hypothesis that fluctuations in the signature of the spacetime
metric are suppressed. In light of the fact that even this `well-behaved'
signature change system predicts its own downfall, it may be prudent to
reassess the inclusion of signature changing metrics in quantum gravity
theories.

{\Large\bf Acknowledgements}

The authors are grateful for helpful discussions with Innigo Egusquiza, Gary
Gibbons, Bernard Kay and Robin Tucker. CJF thanks Churchill College, Cambridge,
the Royal Society and the Schweizerischer Nationalfonds for financial support.

\newpage
\section*{Appendix A}
\setcounter{section}{1}
\renewcommand{\thesection}{\Alph{section}}
\setcounter{equation}{0}

The general form of the boundary conditions for the Schr\"odinger
Hamiltonian are given by
equation (\ref{generalBC}).
Let $A$ and $B$ be the matrices defined by (\ref{AB}), then we are
interested in the subset of boundary conditions given by the restriction
that both $A$ and $B$ are singular. For such $A$ and $B$, the
of the boundary conditions are of the form
\begin{eqnarray}
\omega_1\:\rho^\dagger (0^\tm) &=& \omega_2\:\rho^\dagger (0^\tp) \; ,
\nonumber \\
&& \\
\omega_3\:\frac{\dd \rho^\dagger}{\dd z}(0^\tm) &=&
\omega_4\: \frac{\dd \rho^\dagger}{\dd z}(0^\tp) \; ,
\nonumber
\end{eqnarray}
for some $\omega_1,\omega_2,\omega_3,\omega_4\in\complex$. We also wish to
exclude any cases for which any of the $\omega_i$ are zero.

We can calculate the general
form of the unitary matrix $U$ which satisfies $\det(A)=0$, $\det(B)=0$,
and the constraints (\ref{conditions}) on $a$, $b$, $c$ and $d$.
We find the two parameter family
\begin{equation}
U = \left(
\begin{array}{cc}
-\frac{1}{2} (1-i) (1+i{\rm e}^{i\theta})
& \sqrt{\sin(\theta)}\: {\rm e}^{i\kappa} \\
& \\
-i\sqrt{\sin(\theta)}\: {\rm e}^{i(\theta-\kappa)}
& -\frac{1}{2} (1+i) (1+i{\rm e}^{i\theta})
\end{array} \right)
\end{equation}
for any $\theta\in[0,\pi]$ and $\kappa\in[0,2\pi]$.

Except for two special cases (when $\theta=0$ and $\theta=\pi$), the
boundary conditions corresponding to this $U$ are
\begin{eqnarray}
\rho^\dagger (0^\tm) &=& \left(
\frac{1+\frac{i}{2} (1+i) (1+i{\rm e}^{i\theta})}{\sqrt{\sin(\theta)}\:
{\rm e}^{i\kappa}}
\right)\:\rho^\dagger (0^\tp) \; , \label{appgbc1} \\
&& \nonumber \\
\frac{\dd \rho^\dagger}{\dd z}(0^\tm) &=& \left(
\frac{\frac{1}{2} (1+i) (1+i{\rm e}^{i\theta})-1}{\sqrt{\sin(\theta)}\:
{\rm e}^{i\kappa}}
\right)\: \frac{\dd \rho^\dagger}{\dd z}(0^\tp) \label{appgbc2}
\end{eqnarray}

In the special case $\theta = 0$, the boundary conditions are
\begin{eqnarray}
\rho^\dagger (0^\tm) &=& 0 \; , \nonumber\\
&& \\
\frac{\dd \rho^\dagger}{\dd z}(0^\tp) &=& 0 \; , \nonumber
\end{eqnarray}
and the special case $\theta = \pi$ gives the boundary conditions
\begin{eqnarray}
\rho^\dagger (0^\tp) &=& 0 \; , \nonumber\\
&& \\
\frac{\dd \rho^\dagger}{\dd z}(0^\tm) &=& 0 \; . \nonumber
\end{eqnarray}
These special cases correspond to setting $\omega_2=\omega_3=0$
and $\omega_1=\omega_4=0$ respectively, and hence
we will ignore them.

It seems reasonable to assume that the matter
field remains continuous across
the surface of signature change. In this case, using (\ref{appgbc1})
to eliminate $\kappa$ from (\ref{appgbc2}) we find that the
derivative boundary condition becomes
\begin{equation}
\frac{\dd \rho^\dagger}{\dd z}(0^\tm) =
\left(\frac{\sin(\theta)}{\cos(\theta)-1}\right) \:
\frac{\dd \rho^\dagger}{\dd z}(0^\tp)
\label{derivBC}
\end{equation}

Finally, we can employ something particular to this signature
change system. We want every mode solution to satisfy the boundary conditions.
Furthermore, for $E>0$ and $z>0$ the only bounded solution of
${H}\rho^\dagger = E\rho^\dagger$
is
\begin{equation}
\rho^\dagger(z) = \rho^\dagger (0^\tp) {\rm e}^{-\sqrt{E}\:z}
\end{equation}
and hence we have the additional information that
\begin{equation}
\frac{\dd \rho^\dagger}{\dd z}(0^\tp) = -\sqrt{E} \rho^\dagger (0^\tp) \; .
\label{extrainfo}
\end{equation}

Using equations (\ref{derivBC}) and (\ref{extrainfo}),
equation (\ref{generalBC}) may then be rewritten as
\begin{equation}
-\sqrt{E} A \left(\begin{array}{c}
\left(\frac{\sin(\theta)}{\cos(\theta)-1}\right) \\
\\
1
\end{array} \right)
= {\rm e}^{i\pi/4} B \left(\begin{array}{c} 1 \\ \\ 1 \end{array} \right) \; ,
\end{equation}
but this must hold for all $E>0$ and so we must have
\begin{equation}
A \left(\begin{array}{c}
\left(\frac{\sin(\theta)}{\cos(\theta)-1}\right) \\
\\
1
\end{array} \right)
= \left(\begin{array}{c} 0 \\ \\ 0 \end{array} \right)
\;\;\;\;\;\;{\rm and}\;\;\;\;\;\;
B \left(\begin{array}{c} 1 \\ \\ 1 \end{array} \right)
= \left(\begin{array}{c} 0 \\ \\ 0 \end{array} \right)
\end{equation}
These equations are consistent only if $\theta = \frac{\pi}{2}$.

Thus, setting $\theta = \frac{\pi}{2}$ in (\ref{derivBC}),
we have shown that the only set of boundary conditions of the form
(\ref{niceBC}), which provide a self-adjoint extension of the
Schr\"odinger operator on the signature changing spacetime (\ref{dcmetric}),
are
\begin{eqnarray}
\rho^\dagger (0^\tm) &=& \rho^\dagger (0^\tp) \; , \nonumber \\
&& \\
\frac{\dd \rho^\dagger}{\dd z}(0^\tm) &=& -\:
\frac{\dd \rho^\dagger}{\dd z}(0^\tp) \; .
\nonumber
\end{eqnarray}

\newpage
\section*{Appendix B}
\setcounter{section}{2}
\renewcommand{\thesection}{\Alph{section}}
\setcounter{equation}{0}

Here, we prove that the integral transform $\cal M$ is unitary. It is
convenient to work with the operator $\widetilde{\cal M}$ defined by
\begin{equation}
\widetilde{\cal M} = V {\cal M} U^*,
\end{equation}
where $U:L^2(\real,dE)\rightarrow L^2(\real^+,dk)\otimes\complex^2$ and
$V:L^2(\real,dz)\rightarrow L^2(\real^+,dz)\otimes\complex^2$ are unitary
operators given by
\begin{equation}
(Uf)(k) = \left(\begin{array}{c} (2k)^{1/2} f(k^2) \\ (2k)^{1/2} f(-k^2)
\end{array}\right) \quad{\rm and}\quad
(Vf)(z) = \left(\begin{array}{c} f(z) \\  f(-z)
\end{array}\right) \; .
\end{equation}
Explicitly, $\widetilde{\cal M}:L^2(\real^+,dk)\otimes\complex^2
\rightarrow L^2(\real^+,dz)\otimes\complex^2$ takes the matrix form
\begin{equation}
\widetilde{\cal M} = \left(\begin{array}{cc}
(1+i)A & (1-i)B \\ (1+i)B & (1-i)A \end{array}\right) \; ,
\end{equation}
where $A,B:L^2(\real^+,dk)\rightarrow L^2(\real^+,dz)$ are defined by
\begin{equation}
(Af)(z) = \frac{1}{(2\pi)^{1/2}}\int_0^\infty e^{-kz}f(k) dk
\end{equation}
and
\begin{equation}
(Bf)(z) = \frac{1}{2} (({\cal C}-{\cal S})f)(z) \; ,
\end{equation}
and ${\cal C}$ and ${\cal S}$ are the cosine and sine transforms defined by
\begin{equation}
({\cal C}f)(z) =\left(\frac{2}{\pi}\right)^{1/2}
\int_0^\infty dk f(k) \cos kz
\end{equation}
and
\begin{equation}
({\cal S}f)(z) =\left(\frac{2}{\pi}\right)^{1/2}
\int_0^\infty dk  f(k)\sin kz \; .
\end{equation}

In order to prove that $\widetilde{\cal M}$ (and hence $\cal M$) is unitary,
it
now suffices to establish the following identities:
\begin{eqnarray}
2(A^*A + B^*B) = \id & & 2(AA^*+BB^*) = \id  \nonumber \\
A^*B + B^*A =0 && AB^* + BA^* =0 \; .
\end{eqnarray}
Let $f\in C_0^\infty(0,\infty)$. Then
\begin{eqnarray}
(A^*Bf)(k)&=&\frac{1}{2\pi}\int_0^\infty dz e^{-kz}
\int_0^\infty dk^\prime (\cos k^\prime z-\sin k^\prime z) f(k^\prime)
\nonumber \\
& = & \frac{1}{2\pi}
\int_0^\infty dk^\prime f(k^\prime) G(k,k^\prime) \; ,
\end{eqnarray}
where
\begin{equation}
G(k,k^\prime) = \int_0^\infty dz e^{-kz} (\cos k^\prime z-\sin k^\prime z)
=\frac{k-k^\prime}{k^2+{k^\prime}^2} \; ,
\end{equation}
and Fubini's theorem \cite{R&SI} has been employed. Moreover, an analogous
argument shows that
\begin{equation}
(B^*Af)(k)=\frac{1}{2\pi}
\int_0^\infty dk^\prime f(k^\prime) G(k^\prime,k) \; .
\end{equation}
Thus, because $G(k,k^\prime)=-G(k^\prime,k)$, we conclude that $A^*B+B^*A$
vanishes on $C_0^\infty(0,\infty)$ and hence on the whole of
$L^2(\real^+,dk)$.
An identical argument (with $z$ and $k$ interchanged)
shows that $AB^*+BA^*=0$.

Next, note that $2B^*B = \id -
\frac{1}{2}({\cal S}^*{\cal C} + {\cal C}^*{\cal S})$. Thus we need to
show that
$\frac{1}{2}({\cal S}^*{\cal C} + {\cal C}^*{\cal S}) = 2A^*A$.
Let $f\in C_0^\infty(0,\infty)$, then Fubini's theorem may be used to
show that
\begin{equation}
2(A^*A f)(k) = \frac{1}{\pi} \int_0^\infty
dk^\prime\frac{f(k^\prime)}{k+k^\prime}\; ,
\end{equation}
and also that
\begin{eqnarray}
\mbox{$\frac{1}{2}$}(({\cal S}^* e^{-\epsilon z}{\cal C} +
{\cal C}^* e^{-\epsilon z}{\cal S}))f(k) &=&
\frac{1}{\pi}\int_0^\infty dk^\prime
f(k^\prime)\frac{k+k^\prime}{\epsilon^2+(k+k^\prime)^2} \nonumber \\
&\stackrel{L^2}{\longrightarrow}&
\frac{1}{\pi}\int_0^\infty dk^\prime
\frac{f(k^\prime)}{k+k^\prime}
\end{eqnarray}
as $\epsilon\rightarrow 0^+$. Using the fact that ${\cal S}^*
e^{-\epsilon z}{\cal C} +
{\cal C}^* e^{-\epsilon z}{\cal S} \rightarrow {\cal S}^*{\cal C} + {\cal
C}^*{\cal S}$ strongly as $\epsilon\rightarrow 0^+$, we have $2(A^*A+B^*B)
=\id$ on $C_0^\infty(0,\infty)$ and hence on $L^2(\real^+,dk)$. An analogous
argument shows that $2(AA^*+BB^*)=\id$.



\newpage
\begin{thebibliography}{99}

\bibitem{GibbHart} G. W. Gibbons and J. B. Hartle,
{\it Real Tunneling Geometries and
the Large-Scale Topology of the Universe},
Phys. Rev. D {\bf 42}, 2458--2468 (1990).

\bibitem{TGRG} T. Dray, C. A. Manogue and R. W. Tucker,
{\it Particle Production from Signature Change},
Gen. Rel. Grav. {\bf 23}, 967--971 (1991).

\bibitem{Ellis} G. Ellis, A. Sumeruk, D. Coule and C. Hellaby,
{\it Change of Signature in Classical Relativity},
Class. Quantum. Grav. {\bf 9}, 1535--1554 (1992).

\bibitem{Hayward} S. A. Hayward, {\it Signature Change in General Relativity},
Class. Quantum Grav. {\bf 9}, 1851--1862 (1992).

\bibitem{Hay2} S. A. Hayward, {\it Junction Conditions for Signature Change},
Max Planck Institute preprint 93--0451 (1993).

\bibitem{Tucker} T. Dray, C. A. Manogue and R. W. Tucker, {\it Scalar Field
Equation in the Presence of Signature Change},
Phys. Rev. D {\bf 48}, 2587--2590 (1993).

\bibitem{KK} M. Kossowski and M. Kriele, {\it Signature Type Change and
Absolute Time in General Relativity}, Class. Quantum Grav. {\bf 10},
1157--1164 (1993).

\bibitem{KK2} M. Kossowski and M. Kriele, {\it Smooth and Discontinuous
Signature Type Change in General Relativity}, Class. Quantum Grav. {\bf 10},
2363--2371 (1993).

\bibitem{KK3} M. Kossowski and M. Kriele, {\it The Einstein Equation for
Signature Type Changing Spacetimes}, Proc. Roy. Soc. Lond. A{\bf 446},
115--126 (1994).

\bibitem{Alty} L. J. Alty, {\it Kleinian Signature Change},
Class. Quantum Grav. {\bf 11}, 2523--2536 (1994).

\bibitem{R&SII} M. Reed and B. Simon, {\it Methods of Modern
Mathematical Physics II: Fourier Analysis, Self-Adjointness},
Academic Press (1975).

\bibitem{Bognar} J. Bogn\'ar, {\it Indefinite Inner Product Spaces},
Springer--Verlag (1974).

\bibitem{Thaller} B. Thaller, {\it The Dirac Equation},
Springer--Verlag (1992).

\bibitem{Seba} P. \v{S}eba, {\it The Generalized Point Interaction in One
Dimension},  Czech. J. Phys. B {\bf 36}, 667--673 (1986).

\bibitem{Bernard} B. S. Kay and U. M. Studer, {\it Boundary Conditions
for Quantum Mechanics on Cones and Fields around Cosmic Strings},
Commun. Math. Phys. {\bf 139}, 103--139 (1991).

\bibitem{R&SI} M. Reed and B. Simon, {\it Methods of Modern
Mathematical Physics I: Functional Analysis}, Academic Press (1972).

\bibitem{C+H} R. Courant and D. Hilbert, {\it Methods of Mathematical
Physics: Volume~II}, Interscience Publishers (1962).

\bibitem{Innigo} I. L. Egusquiza, {\it Self-adjoint Extensions and Signature
Change}, Preprint gr-qc/9503015.

\end{thebibliography}
\end{document}